\documentclass[reprint,amsmath,amssymb,aps,prl]{revtex4-1}

\usepackage{amsmath}    
\usepackage{graphics}
\usepackage{graphicx}   
\usepackage{dcolumn}    
\usepackage{bm}         
\usepackage{verbatim}   
\usepackage{color}
\definecolor{linkc}{RGB}{51,51,153}
\usepackage[colorlinks=true,allcolors=linkc]{hyperref}

\begin{document}

\title{Quantum nondemolition measurement of mechanical squeezed state beyond the 3 dB limit}

\author{C. U. Lei,$^1$ A. J. Weinstein,$^1$ J. Suh,$^2$ E. E. Wollman,$^1$ A. Kronwald,$^{3,4}$ F. Marquardt,$^{3,4}$ A. A. Clerk,$^{5}$ K. C. Schwab$^{1}$}
\email{schwab@caltech.edu}
\affiliation{$^{1}$Applied Physics, California Institute of Technology, Pasadena, CA 91125, USA}
\affiliation{$^{2}$Korea Research Institute of Standards and Science, Daejeon 305-340, Republic of Korea}
\affiliation{$^{3}$Friedrich-Alexander-Universit{\"a}t Erlangen-N{\"u}rnberg, Staudtstr. 7, D-91058 Erlangen, Germany}
\affiliation{$^{4}$Max Planck Institute for the Science of Light G{\"u}nther-Scharowsky-Stra{\ss}e 1/Bau 24, D-91058 Erlangen, Germany}
\affiliation{$^{5}$Department of Physics, McGill University, Montreal, Quebec, H3A 2T8 Canada}

\date{\today}

\begin{abstract}
We use a reservoir engineering technique based on two-tone driving to generate and stabilize a quantum squeezed state of a micron-scale mechanical oscillator in a microwave optomechanical system. Using an independent backaction evading measurement to directly quantify the squeezing, we observe $4.7\pm 0.9$ dB of squeezing below the zero-point level, surpassing the 3 dB limit of standard parametric squeezing techniques. Our measurements also reveal evidence for an additional mechanical parametric effect.  The interplay between this effect and the optomechanical interaction enhances the amount of squeezing obtained in the experiment.
\end{abstract}

\maketitle

Generating nonclassical states of a massive object has been a subject of considerable interest. It offers a route toward fundamental tests of quantum mechanics in an unexplored regime \cite{aspelmeyer2012}. One of the most important and elementary quantum states of an oscillator is a squeezed state \cite{stoler1970equivalence}:  a minimum uncertainty state has a quadrature which is smaller than the zero-point level. Such states have long been discussed in the context of gravitational waves detection to improve the measurement sensitivity \cite{hollenhorst1979quantum,caves1981quantum}. It is well known that a coherent parametric drive can be used to squeeze mechanical fluctuations \cite{blencowe2000quantum,rugar1991mechanical}, which is essentially equivalent to the technique first used to squeeze ground-state optical fields \cite{wu1986generation}. However, the maximum steady-state squeezing achieved by this method is limited to 3 dB due to the onset of parametric instability. Therefore, it is in principle impossible to have a steady state where the mechanical motion is squeezed below one half of the zero-point level using only parametric driving. These limitations may be overcome by combining continuous quantum measurement and  feedback \cite{ruskov2005squeezing,clerk2008back,szorkovszky2011mechanical,szorkovszky2013strong}, but it would substantially increase the experimental complexity.

Another method to generate robust quantum state is quantum reservoir engineering \cite{verstraete2009quantum}, which has been used to generate quantum squeezed states and entanglement with trapped ions \cite{lin2013dissipative, kienzler2015quantum} and superconducting qubits \cite{shankar2013autonomously}. It can also applied to optomechanical system to generate strong steady-state squeezing without quantum-limited measurement and feedback \cite{kronwald2013arbitrarily}. By modulating the optomechanical coupling with two imbalanced classical drive tones, the driven cavity acts effectively as a squeezed reservoir. When the engineered dissipation from the cavity dominates the dissipation from the environment, the mechanical resonator relaxes to a steady squeezed state. This technique has been applied recently to generate quantum squeezed states of macroscopic mechanical resonators \cite{wollman2015quantum, pirkkalainen2015squeezing,lecocq2015quantum}.

In addition to being a tool for state preparation, optomechanics also provides a means to probe the quantum behavior of macroscopic objects \cite{weinstein2014observation, cohen2015phonon, lecocq2015resolving}. In particular, a backaction evading (BAE) measurement \cite{braginsky1980quantum, braginsky1996quantum, clerk2008back, hertzberg2010back, suh2014mechanically, lecocq2015quantum} of a single motional quadrature can be implemented in an optomechanical system. If the drive tones that modulate the coupling are balanced, a continuous quantum nondemolition (QND) measurement of the mechanical quadrature can be made. This technique can be used to fully reconstruct the quantum state of the mechanical motion.

In this work, we combine reservoir engineering and backaction evading measurement with a microwave optomechanical system to perform continuous QND measurement of a quantum squeezed state. Among the previous three squeezing experiments \cite{wollman2015quantum, pirkkalainen2015squeezing,lecocq2015quantum}, only \cite{lecocq2015quantum} demonstrated direct detection, performed using a two-cavity optomechanical system; here we implement both reservoir engineering and BAE measurement simultaneously within a simple single-cavity setup. In addition to the optomechanical interaction, a mechanical parametric effect is observed. Contrary to previous works, where the mechanical parametric effect produced parametric instability that limited the precision of the BAE measurement \cite{hertzberg2010back,suh2013optomechanical,suh2012thermally}, the interplay between the parametric drive and the engineered dissipation enhances the mechanical squeezing. By directly measuring the mechanical quadrature variances with the BAE measurement, we demonstrate motional quantum squeezing with squeezed quadrature variance $\langle\Delta X_1^2\rangle = 0.34\pm0.07 x_{\textrm{zp}}^2$, $4.7\pm0.9$ dB below the zero-point level. This exceeds what is possible using only parametric driving, even if one starts in the quantum ground state. This is the first experiment to demonstrate more than 3 dB quantum squeezing in a macroscopic mechanical system.

\begin{figure*}[p]
\begin{center}
\includegraphics{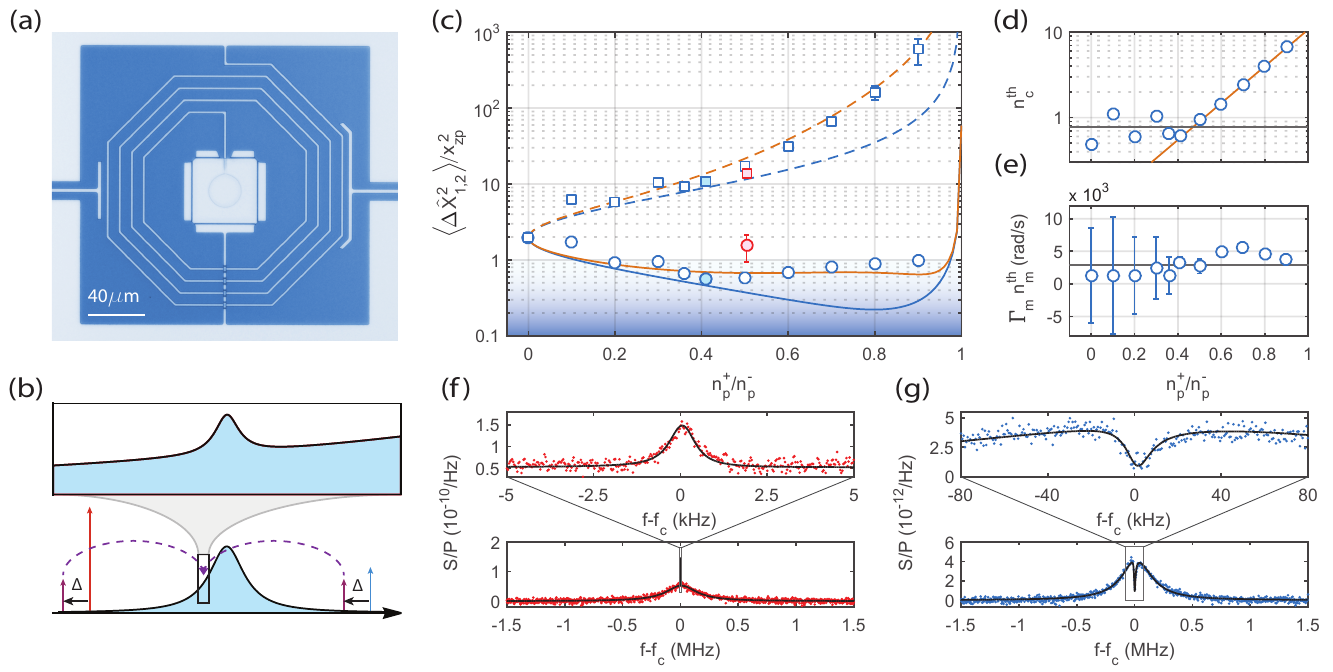}
\end{center}
\caption{(a) Optical micrograph of the device. The gray region is aluminum, the blue region is silicon. The square at the center is a parallel plate capacitor which is coupled to a spiral inductor to form a microwave resonator. The top gate of the capacitor is a compliant membrane whose fundamental motion is being studied. (b) Schematic of the pumps (red and blue arrows)
and probes (purple arrows) relative to the cavity resonance. The inset shows the schematic of the BAE probe sideband spectrum. (c) The squeezed quadrature variances (circles) and anti-squeezed quadrature variances (squares) inferred from the output spectra. The red (blue) symbols represent the squeezed states achieved at $n_p^{\textrm{tot}} = 1.35\times10^4$ ($n_p^{\textrm{tot}} = 1.85\times10^5$). The blue shaded region indicates sub-zero point squeezing. The blue curves are the predictions from Eq. (\ref{eq:X12sq}) with constant cavity and mechanical occupations at $n_p^+/n_p^-=0$. The orange curves are the predictions from Eq. (\ref{eq:X12sq}) including cavity heating effect extracted from the experiment. (d) The cavity occupation $n_c^{\textrm{th}}$ extracted from the output spectrums, the orange line is a linear fit of the pump ratio dependent heating. (e) The phonon bath heating rate $\Gamma_m n_m^{\textrm{th}}$ extracted from the output spectrums. (f) (g) The output spectra normalized by the transmitted power of the red-detuned pump. (f) The normalized output spectra correspond to the solid red circle in Fig.~\ref{fig:1}c. (g) The normalized output spectra correspond to the solid blue circle in Fig.~\ref{fig:1}c.}
\label{fig:1}
\end{figure*}

The mechanical oscillator in this work is a 100 nm thick, $40\times40\,\mu\textrm{m}^2$ aluminum membrane, with fundamental resonance frequency $\omega_m=2\pi\times5.8$ MHz and mechanical linewidth $\gamma_m=2\pi\times8$ Hz at 10 mK. It is capacitively coupled to a lumped-element superconducting microwave resonator with resonance frequency $\omega_c=2\pi\times6.083$ GHz and damping rate $\kappa=2\pi\times330$ kHz (Fig.~\ref{fig:1}a). The mechanical motion couples to the resonance frequency of the microwave resonator through the modulation of the capacitance, with an optomechanical coupling rate $g_0=\frac{d\omega_c}{dx}x_{\textrm{zp}}=2\pi\times130$ Hz, where $x_{\textrm{zp}}=\sqrt{\frac{\hbar}{2m\omega_m}}=1.8$ fm is the amplitude of the zero-point fluctuation of the mechanical oscillator with mass $m=432$ pg. The system is described by the Hamiltonian

\begin{align}
\hat{H} =\ & \hbar\omega_{c}\hat{a}^{\dagger}\hat{a}+\hbar\omega_{m}\hat{b}^{\dagger}\hat{b}-\hbar g_{0}\hat{a}^{\dagger}\hat{a}\left(\hat{b}+\hat{b}^{\dagger}\right) \nonumber \\ &+i\hbar\sqrt{\kappa_{\textrm{in}}}(\alpha^{*}(t)\hat{a}-\alpha(t)\hat{a}^{\dagger})+\hat{H}_{\textrm{diss}},
\label{eq:Hamiltonian}
\end{align}
where $\hat{a}\left(\hat{a}^{\dagger}\right)$ is the annihilation (creation) operator of the intra-cavity field, $\hat{b}\left(\hat{b}^{\dagger}\right)$ is the mechanical phonon annihilation (creation) operator, $\kappa_{\textrm{in}}$ is the coupling rate of the input coupler, and $\alpha(t)$ is the external driving field. The term $\hat{H}_{\textrm{diss}}$ accounts for dissipation.

To squeeze the mechanical motion, we drive the cavity with a pair of pump tones at $\omega_c\mp\omega_m$ with intracavity field \cite{kronwald2013arbitrarily}
\begin{equation}
\bar{\alpha}_{\textrm{sqz}}(t)=(\bar{\alpha}_- e^{i\omega_m t}+\bar{\alpha}_+ e^{-i\omega_m t})e^{-i\omega_c t},
\end{equation}
which is represented by the red and blue arrows in Fig.~\ref{fig:1}b. Linearizing the cavity dynamics in the standard way, the pumps couple the microwave resonator to the Bogoliubov mode of the mechanical motion with the Hamiltonian
\begin{equation}
\hat{H}_{\textrm{sqz}}=-\hbar\ensuremath{\mathcal{G}}(\hat{d}^{\dagger}\hat{\beta}+\hat{d}\hat{\beta}^{\dagger}),
\label{eq:linearHamiltonian}
\end{equation}
where $\hat{d}$ is the fluctuating part of the cavity field $\hat{a}$, and $\hat{\beta}=\hat{b}\cosh{r}+\hat{b}^\dagger\sinh{r}$ is the Bogoliubov-mode annihilation operator whose ground state is a squeezed state with squeezing parameter $r=\tanh^{-1}(G_+/G_-)$. $\ensuremath{\mathcal{G}}=\sqrt{G_-^2-G_+^2}$ is the coupling rate between the Bogoliubov mode and the cavity. $G_\mp=g_0\sqrt{n_p^{\mp}}$ are the enhanced optomechanical coupling rates, and $n_p^{\mp}=|\bar{\alpha}_{\mp}|^2$ are the intracavity pump photon numbers corresponding to the squeezing pumps.

The beam-splitter Hamiltonian in Eq. (\ref{eq:linearHamiltonian}) enables us to cool the Bogoliubov-mode into its ground state, producing a stationary mechanical squeezed state with quadrature variances
\begin{equation}
\langle\Delta\hat{X}_{1,2}^2\rangle = x_{zp}^2 \bigg\{ \frac{\Gamma_m}{\Gamma_{\textrm{eff}}} \left(2 n_m^{th}+1\right) + \frac{\Gamma^{\mp}_{\textrm{opt}}}{\Gamma_{\textrm{eff}}} \left(2 n_c^{th}+1\right) \bigg\},
\label{eq:X12sq}
\end{equation}
where $\Gamma_{\textrm{eff}}=\Gamma_m+4\mathcal{G}^2/\kappa$ is the effective mechanical linewidth and $\Gamma^{\mp}_{\textrm{opt}}=4(G_-\mp G_+)^2/\kappa$ parameterizes the phase-dependent driving of the mechanics by cavity fluctuations. The quadrature variances depend on the intracavity pump photon numbers $n_p^{\mp}$, as well as the cavity occupation $n_c^{\textrm{th}}$ and the phonon bath occupation $n_m^{\textrm{th}}$, which can be extracted from the output spectra (Fig.~\ref{fig:1}f,g). Together with Eq. (\ref{eq:X12sq}), the corresponding quadrature variances can be calculated \cite{kronwald2013arbitrarily,supp2016}.

Fig.~\ref{fig:1}c shows the quadrature variances with various intracavity pump photon ratio $n_p^+/n_p^-$. We start by squeezing the mechanical motion with total intracavity pump photon number $n_p^{\textrm{tot}}=n_p^-+n_p^+=1.35\times10^4$ and pump photon ratio $n_p^+/n_p^-=0.5$. This pump configuration generates a mechanical squeezed state with the squeezed quadrature variance $\langle\Delta\hat{X}_1^2\rangle = 1.54\pm0.59 x_{\textrm{zp}}^2$ and the anti-squeezed quadrature variance $\langle\Delta\hat{X}_2^2\rangle = 13.81\pm1.41 x_{\textrm{zp}}^2$, indicated by the solid red circle and square in Fig.~\ref{fig:1}c. The corresponding normalized output spectra and the fits from the two-tone optomechanical model \cite{kronwald2013arbitrarily} are shown in Fig.~\ref{fig:1}f. To further squeeze the mechanical motion, we can increase the total pump photon number. The blue circles (squares) in Fig.~\ref{fig:1}c are the squeezed (anti-squeezed) quadrature variances at total intracavity pump photon number $n_p^{\textrm{tot}}=1.85\times10^5$. The solid (dashed) blue curves are the predictions from Eq. (\ref{eq:X12sq}) with constant cavity and mechanical occupations extracted from the output spectrum at $n_p^+=0$; they agree with the data at low pump photon ratio. At large pump photon ratio, the cavity bath starts to heat up (Fig.~\ref{fig:1}d), which increases the mechanical quadrature variances. The orange curves in Fig.~\ref{fig:1}c are the predictions from Eq. (\ref{eq:X12sq}) including the cavity heating effect extracted from the experiment (orange line in Fig.~\ref{fig:1}d). With the heating effect, the minimum quadrature variance is achieved at $n_p^+/n_p^-=0.43$ with $\langle\Delta\hat{X}_1^2\rangle = 0.56\pm0.02 x_{\textrm{zp}}^2$ (the solid blue circle in Fig.~\ref{fig:1}d), $2.5\pm0.2$ dB below the zero-point level. The corresponding normalized output spectra and fits are shown in Fig.~\ref{fig:1}g.

While inferring the level of squeezing from the cavity output spectrum is convenient, it would be preferable to have a more direct method that does not rely on assumptions about the mechanical dynamics.  This can be achieved in our system without needing to introduce an additional cavity resonance:  we continue to use the cavity density of states near resonances to generate mechanical squeezing, but now use the density of states away from resonances to make an independent, backaction-evading measurement of a single mechanical quadrature. In this way, our single cavity effectively plays the role of two:  it both generates squeezing, and permits an independent detection of the squeezing.

\begin{figure*}[p]
\begin{center}
\includegraphics{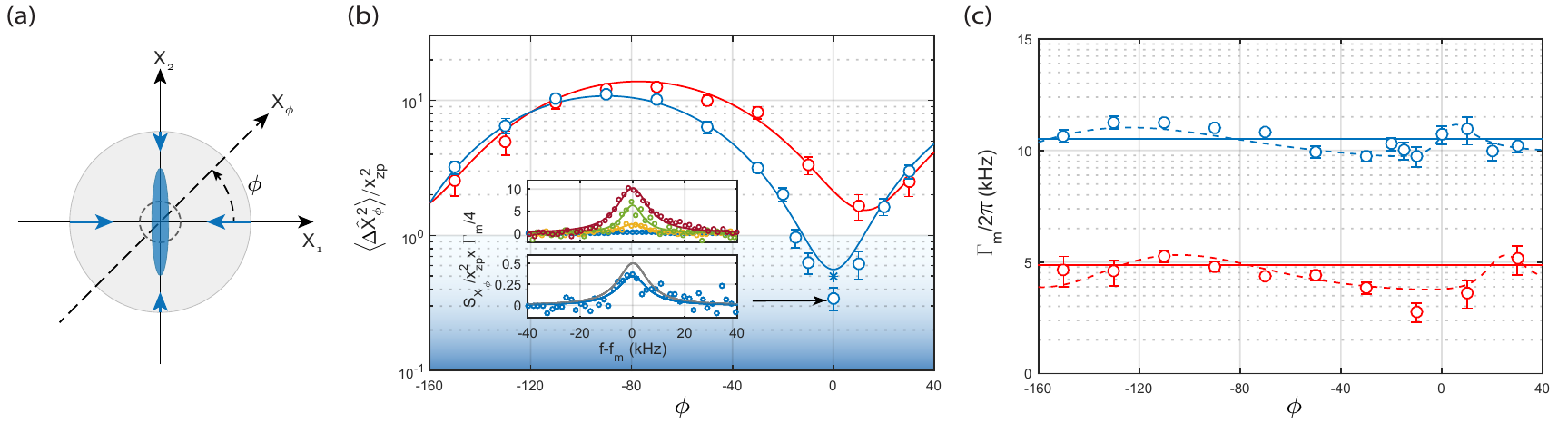}
\end{center}
\caption{(a) Schematic of dissipative mechanical squeezing. The gray circle represents the initial thermal state in phase space. The engineered reservoir generates phase dependent dissipation that relaxes the mechanics into a squeezed state, which is represented by the blue ellipse. The gray dashed circle represents the zero-point level. (b) Mechanical quadrature variance as a function of probe phase. The blue shaded region indicates sub-zero point squeezing. The red (blue) circles are the quadrature variances of the weakly (strong) squeezed state as measured using the BAE technique. The red (blue) curves are the quadrature variances inferred from the corresponding output spectra assuming no mechanical parametric drive. The deviation of the blue curve and circle at $\phi=0^\circ$ indicates the importance of the mechanical parametric drive. An optomechanical model including this effect explains the extra squeezing (blue star at $\phi=0^\circ$). The insets are the mechanical quadrature spectra of the strong squeezed state with phase $\phi$ at $-70^{\circ}$ (red), $-50^{\circ}$ (green), $-20^{\circ}$ (yellow), $0^{\circ}$ (blue). The gray Lorentzian in the lower inset represents the spectrum with quadrature variance equal to half of the zero-point fluctuation (the 3 dB limit). (c) Mechanical quadrature linewidth as a function of probe phase. The red (blue) circles are the measured mechanical quadrature linewidth of the weakly (strong) squeezed state. The solid lines are the theoretical predictions from the ideal optomechanical model. The dashed curves are the fit with the optomechanical model including the mechanical parametric interaction.}
\label{fig:2}
\end{figure*}

To directly measure a single mechanical quadrature, in addition to the squeezing pumps, we introduce another pair of weak backaction evading (BAE) probes (the purple arrows in Fig.~\ref{fig:1}b) at $\omega_c\mp\omega_m-\Delta$ with intracavity field \cite{suh2014mechanically,wollman2015quantum}
\begin{equation}
\bar{\alpha}_{\textrm{BAE}}(t)=2\bar{\alpha}\cos(\omega_m t+\phi)e^{-i(\omega_c-\Delta)t},
\end{equation}
where $\phi$ is the relative phase between the BAE probes and the squeezing pumps. For a sideband-resolved system ($\omega_m\gg\kappa$), the modulation of the BAE probes exclusively couples the mechanical quadrature $\hat{X}_{\phi}=\cos{\phi}\hat{X}_1-\sin{\phi}\hat{X}_2$ to the microwave resonance with the interaction
\begin{equation}
\hat{H}_{I}=-\hbar G(\hat{d}^{\dagger}e^{-i\Delta t}+\hat{d}e^{i\Delta t})\frac{\hat{X}_{\phi}}{x_{\textrm{zp}}},
\label{eq:InteractionHamiltonian}
\end{equation}
where $G=g_0\sqrt{n_p}$ is the enhanced optomechanical coupling rate and $n_p=|\bar{\alpha}|^2$ is the intracavity pump photon number corresponding to the BAE probes. Since $\hat{X}_{\phi}$ is a constant of motion of the system, the interaction (\ref{eq:InteractionHamiltonian}) enables a continuous QND measurement of the the mechanical quadrature. By sweeping the probe phase $\phi$, we can perform tomography of the mechanical quantum state (Fig. ~\ref{fig:2}a). In order to ensure no interference between the sidebands of the squeezing pumps and the BAE probes, we detune the BAE sidebands from the cavity resonance by $\Delta = 2\pi\times160\textrm{kHz}\gg \Gamma_{\textrm{eff}}$. The power of the BAE probes are set about 10 dB weaker than the power of the squeezing pumps to avoid extra heating. In the experiment, the motional sideband spectrum of the BAE probes is measured, from which we can extract the mechanical quadrature variance and linewidth. In the following, we will perform a BAE measurement to directly characterize the weakly squeezed state corresponding to the spectrum Fig. ~\ref{fig:1}f and the strong squeezed state corresponding to the spectrum Fig.~\ref{fig:1}g.

Fig.~\ref{fig:2}b shows the mechanical quadrature variances from the BAE measurement as a function of the probe phase $\phi$. The red circles are the quadrature variances of the weakly squeezed state measured with the BAE technique. The red curve is the inferred quadrature variances from the corresponding output spectrum (Fig.~\ref{fig:1}f). In this case, the results from the BAE measurement are in good agreement with the results inferred from the output spectrum. Similarly, the blue circles are the quadrature variances of the strong squeezed state measured with the BAE technique. The blue curve is the inferred quadrature variance from the corresponding output spectrum (Fig. ~\ref{fig:1}g). The minimum quadrature variance is achieved at $\phi=0^\circ$ with $\langle\Delta\hat{X}^2_{\phi} \rangle = 0.34\pm0.07 x_{\textrm{zp}}^2$, $4.7\pm0.9$ dB below the zero-point level. This is lower than the quadrature variance inferred from the output spectrum, implying that there is additional dynamics at play (beyond the ideal optomechanical interaction).

The enhanced squeezing observed in the BAE measurement suggests an additional squeezing mechanism beyond the dissipative mechanism discussed above; an obvious candidate is direct parametric driving of the mechanics.  The presence of such driving is further corroborated by our observation of a phase dependence of the quadrature linewidth in the BAE measurement (Fig.~\ref{fig:2}c). Similar induced mechanical parametric driving has been observed in other BAE measurements; they can arise via a number of mechanisms, including thermal effects as well as higher nonlinearities \cite{suh2013optomechanical,suh2012thermally}. To understand the effects of this mechanical parametric driving, we phenomenologically add the mechanical parametric interaction to our otherwise ideal optomechanical model \cite{supp2016}:
\begin{equation}
\hat H_\text{para} = -\hbar\lambda(e^{i\psi}\hat{b}^2+e^{-i\psi}\hat{b}^{\dagger2}),
\label{eq:Hpara}
\end{equation}
where $\lambda$ is the amplitude of the parametric interaction and $\psi$ is the relative phase between the parametric drive and the squeezing pumps.

We fit the observed phase-dependent quadrature linewidth to our model, thus extracting the amplitude and phase of the parametric drive. \cite{supp2016}. By assuming the phase of the parametric drive $\psi$ follows the phase of the BAE probe (i.e. $\psi=\phi+\psi_0$, where $\psi_0$ is a constant phase shift), the model captures the observed phase dependence behavior of the quadrature linewidth, as shown by the dashed curves in Fig.~\ref{fig:2}c. Surprisingly, if one instead assumes that the parametric driving is a result of the main squeezing tones (i.e. take $\psi$ a constant independent of $\phi$), one cannot capture the observed phase dependence of the quadrature linewidth \cite{supp2016}. These results suggest that the parametric drive is induced by the BAE probes.

The predicted squeezed quadrature variance for the strong-pumps configuration is $\langle\Delta\hat{X}_1^2\rangle = 0.50\pm0.05 x_{\textrm{zp}}^2$, as shown by the blue star in Fig.~\ref{fig:2}b. We stress that our treatment of the spurious mechanical parametric drive is phenomenological; we do not know the precise microscopic mechanism which causes this driving.  Nonetheless, it allows us to explain both surprising features of the BAE measurements (the observed phase-dependent mechanical quadrature linewidth, and the enhanced squeezing).

In conclusion, we combine reservoir engineering and backaction evading measurement in a microwave optomechanical system to demonstrate a continuous QND measurement of a mechanical quantum squeezed state. From the BAE measurement, $4.7\pm0.9$ dB of squeezing below the zero point level has been observed, surpassing the 3 dB limit of the standard parametric squeezing technique. In addition, a phase dependence of the quadrature linewidth is observed and explained by including a mechanical parametric interaction to the ideal optomechanical model. The interplay between the optomechanical interaction and the mechanical parametric interaction enhances the mechanical squeezing and provides a qualitative explanation to the BAE measurement results. The present scheme can be applied to generate and characterize more complicated quantum states by carefully engineering the nonlinear interaction \cite{woolley2013two,woolley2014two}. The ability to generate and measure a strong quantum squeezed state of a macroscopic mechanical object would be useful for ultra-sensitive detection \cite{hollenhorst1979quantum}, quantum information processing \cite{braunstein2005quantum}, as well as fundamental study of quantum decoherence \cite{zurek2003decoherence,hu1993squeezed}.

This work is supported by funding provided by the Institute for Quantum Information and Matter, an NSF Physics Frontiers Center with support of the Gordon and Betty Moore Foundation (NSF-IQIM 1125565), by the Defense Advanced Research Projects Agency (DARPA-QUANTUM HR0011-10-1-0066), by the NSF (NSF-DMR 1052647 and NSF-EEC 0832819), and by the Semiconductor Research Corporation (SRC) and Defense Advanced Research Project Agency (DARPA) through STARnet Center for Function Accelerated nanoMaterial Engineering (FAME). A.C., F.M., and A.K. acknowledge support from the DARPA ORCHID program through a grant from AFOSR, F.M. and A.K. from ITN cQOM and the ERC OPTOMECH, and A.C. from NSERC.

\bibliographystyle{apsrev4-1}
\bibliography{BAEnSqz_2}

\pagebreak
\onecolumngrid
\clearpage

\begin{center}
\textbf{\large Supplementary Information for ``Quantum nondemolition measurement of mechanical squeezed state beyond the 3 dB limit"}
\end{center}
\begin{center}
C. U. Lei,$^1$ A. J. Weinstein,$^1$ J. Suh,$^2$ E. E. Wollman,$^1$ A. Kronwald,$^{3,4}$ F. Marquardt,$^{3,4}$ A. A. Clerk,$^{5}$ K. C. Schwab$^{1*}$
\end{center}
\begin{center}
\text{\small
$^{1}$Applied Physics, California Institute of Technology, Pasadena, CA 91125, USA} \\
\text{\small$^{2}$Korea Research Institute of Standards and Science, Daejeon 305-340, Republic of Korea} \\
\text{\small$^{3}$Friedrich-Alexander-Universit{\"a}t Erlangen-N{\"u}rnberg, Staudtstr. 7, D-91058 Erlangen, Germany} \\
\text{\small$^{4}$Max Planck Institute for the Science of Light G{\"u}nther-Scharowsky-Stra{\ss}e 1/Bau 24, D-91058 Erlangen, Germany} \\
\text{\small$^{5}$Department of Physics, McGill University, Montreal, Quebec, H3A 2T8 Canada}
\end{center}

\setcounter{equation}{0}
\setcounter{figure}{0}
\setcounter{table}{0}
\setcounter{page}{1}
\makeatletter
\renewcommand{\theequation}{S\arabic{equation}}
\renewcommand{\thefigure}{S\arabic{figure}}

\section{Theory}

\subsection{Ideal two tones optomechanical Hamiltonian}
The Hamiltonian of a generic optomechanical system reads
\begin{equation}
\hat{H}=\hbar\omega_{c}\hat{a}^{\dagger}\hat{a}+\hbar\omega_{m}\hat{b}^{\dagger}\hat{b}-\hbar g_{0}\hat{a}^{\dagger}\hat{a}\left(\hat{b}+\hat{b}^{\dagger}\right)+\hat{H}_{\textrm{drive}},
\end{equation}
where $\hat{a}\left(\hat{a}^{\dagger}\right)$ is the annihilation (creation) operator of the intra-cavity field, $\hat{b}\left(\hat{b}^{\dagger}\right)$
is the mechanical phonon annihilation (creation) operator, and $g_{0}$ is the bare optomechanical coupling between the cavity and the mechanical oscillator. $\hat{H}_{\textrm{drive}}$ describes the external driving.

The device studied in this work is a two ports optomechanical system. Microwave tones are applied from the left port, which we designate $\left(L\right)$. In this section, we consider a system driven by two microwave tones.  The drive Hamiltonian reads
\begin{equation}
\hat{H}_{\textrm{drive}}=\hbar\sqrt{\kappa_{\textrm{L}}}\sum_{\nu=\pm}\alpha_{\nu}\left(\hat{a}e^{i\omega_{\nu}t}+\hat{a}^{\dagger}e^{-i\omega_{\nu}t}\right),
\end{equation}
where $\omega_{\pm}=\omega_{c}+\Delta\pm\left(\omega_{m}+\delta\right)$ and $\alpha_\pm$ are the blue and red pump amplitudes at the input port. In the following, we apply the standard linearization -- i.e., we separate the cavity and the mechanical operators, $\hat{a}$ and $\hat{b}$, into a classical part, $\bar a$ or $\bar b$, plus quantum fluctuations, $\hat d$ or $\hat b$. E.g., $\hat{a} \rightarrow \bar{a} + \hat{d}$. In the interaction picture with respect to  $\hat{H}_{0}=\hbar\left(\omega_{c}+\Delta\right)\hat{a}^{\dagger}\hat{a}+\hbar\left(\omega_{m}+\delta\right)\hat{b}^{\dagger}\hat{b}$, we find the \textit{linearized} optomechanical Hamiltonian
\begin{equation}
\hat H = \hat H_\text{RWA} + \hat H _\text{CR} \, .\label{eq:H_lin_full}
\end{equation}
Here,
\begin{equation}
 \hat H_\text{RWA} = -\hbar \Delta \hat d^\dagger \hat d - \hbar \delta \hat b^\dagger \hat b - \hbar \left[  \left( G_+ \hat b^\dagger + G_- \hat b  \right)\hat d^\dagger +\left(  G_+ \hat b + G_- \hat b^\dagger \right) \hat d \right]
\end{equation}
describes the resonant part of the linearized optomechanical interaction whereas
\begin{equation}
\hat H_\text{CR} = -\hbar \left[ G_+ e^{-2i(\omega_m+\delta)t}\hat b + G_- e^{2i(\omega_m+\delta)t}\hat b^\dagger \right]\hat d^\dagger
-\hbar \left[ G_+ e^{2i(\omega_m + \delta)t } \hat b^\dagger + G_- e^{-2i(\omega_m +\delta)t}\hat b \right] \hat d \label{eq:H_CR}
\end{equation}
describes off-resonant optomechanical interactions. Note that $G_\pm = g_0 \bar a_\pm$ describes the driven-enhanced optomechanical coupling. Here, $\bar a_\pm$ is the intracavity microwave amplitude due to the red and blue pumps, and we have assumed $\bar a_\pm\in\mathbb{R}$ for simplicity and without loss of generality. In the following analysis, we consider the good cavity limit $\left(\omega_{m}\gg\kappa\right)$. At this limit, the off-resonant part of the Hamiltonian can be neglected by the rotating wave approximation (RWA).

\subsection{Mechanical parametric modulation}
In addition to the ideal optomechanical interaction, mechanical parametric modulation is observed in the experiment. This spurious mechanical parametric effect can be induced by thermal effects or nonlinearities \cite{suh2013optomechanical,suh2012thermally}. To take this effect into account, we phenomenologically include the mechanical parametric interaction
\begin{equation}
\hat H_\text{para} = -\hbar\lambda(e^{i\psi}\hat{b}^2+e^{-i\psi}\hat{b}^{\dagger2}),
\label{eq:Hpara}
\end{equation}
where $\lambda$ is the amplitude of the parametric interaction, $\psi$ is the relative phase between the parametric drive and the squeezing pump.

\subsection{Quantum Langevin equations}
The linearized quantum Langevin equations read
\begin{align}
\dot{\hat{d}}&=-\left(\frac{\kappa}{2}-i\Delta\right)\hat{d}+i\left(G_{-}\hat{b}+G_{+}\hat{b}^{\dagger}\right)+\sqrt{\kappa}\hat{d}_{\textrm{in}}, \label{eq:QLE_RWA_d}\\
\dot{\hat{b}}&=-\left(\frac{\Gamma_{m}}{2}-i\delta\right)\hat{b}-i2\lambda e^{-i\psi}\hat{b}^{\dagger}+i\left(G_{-}\hat{d}+G_{+}\hat{d}^{\dagger}\right)+\sqrt{\Gamma_{m}}\hat{b}_{\textrm{in}}\, . \label{eq:QLE_RWA_c}
\end{align}
Here, $\hat{d}_{\textrm{in}}=\sum_{\sigma=L,R,I}\sqrt{\frac{\kappa_{\sigma}}{\kappa}}\hat{d}_{\sigma,\textrm{in}}$
is the total input noise of the cavity, where $\hat{d}_{\sigma,\textrm{in}}$ describes the input fluctuations to the cavity from channel $\sigma$ with damping rate $\kappa_{\sigma}$. $\sigma=L$ and $R$ correspond to the left and right microwave cavity ports, while $\sigma=I$ corresponds to internal losses. The noise operator $\hat{c}_{\textrm{in}}$ describes
quantum and thermal noise of the mechanical oscillator with intrinsic damping rate $\Gamma_{m}$. The input field operators satisfy the following commutation relations:
\begin{eqnarray}
\left[\hat{d}_{\sigma,\textrm{in}}\left(t\right),\hat{d}_{\sigma^{\prime},\textrm{in}}^{\dagger}\left(t^{\prime}\right)\right] & = & \delta_{\sigma,\sigma^{\prime}}\delta\left(t-t^{\prime}\right),\\
\left[\hat{b}_{\textrm{in}}\left(t\right),\hat{b}_{\textrm{in}}^{\dagger}\left(t^{\prime}\right)\right] & = & \delta\left(t-t^{\prime}\right),\\
\left\langle \hat{d}_{\sigma^{\prime},\textrm{in}}^{\dagger}\left(t\right)\hat{d}_{\sigma,\textrm{in}}\left(t^{\prime}\right)\right\rangle  & = & n_{\sigma}^{\textrm{th}}\delta_{\sigma,\sigma^{\prime}}\delta\left(t-t^{\prime}\right),\\
\left\langle \hat{b}_{\textrm{in}}^{\dagger}\left(t\right)\hat{b}_{\textrm{in}}\left(t^{\prime}\right)\right\rangle  & = & n_{m}^{\textrm{th}}\delta\left(t-t^{\prime}\right),
\label{eq:comm}
\end{eqnarray}
where $n_{\sigma}^{\textrm{th}}$ is the photon occupation in port
$\sigma$, and $n_{m}^{\textrm{th}}=1/\left[\exp\left(\hbar\omega_{m}/k_{B}T\right)-1\right]$
is the thermal occupation of the mechanical oscillator. The total
occupation of the cavity is the weighted sum of the contributions from
different channels: $n_{c}^{\textrm{th}}=\sum_{\sigma}\frac{\kappa_{\sigma}}{\kappa}n_{\sigma}^{\textrm{th}}$.

\subsection{Optomechanical output spectrum and mechanical spectrum}
In this section, we derive the optomechanical output spectrum and the mechanical quadrature spectrum. For this, we solve the quantum Langevin equations (Eqs.~\ref{eq:QLE_RWA_d},~\ref{eq:QLE_RWA_c}) in Fourier space. It is convenient to define the vectors $\boldsymbol{D}=\left(\hat{d},\hat{d}^{\dagger},\hat{b},\hat{b}^{\dagger}\right)^{T}$,
$\boldsymbol{D}_{\textrm{in}}=\left(\hat{d}_{\textrm{in}},\hat{d}_{\textrm{in}}^{\dagger},\hat{b}_{\textrm{in}},\hat{b}_{\textrm{in}}^{\dagger}\right)^{T}$and
$\boldsymbol{L}=\textrm{diag}\left(\sqrt{\kappa},\sqrt{\kappa},\sqrt{\Gamma_{m}},\sqrt{\Gamma_{m}}\right)$. We then find the following solution to the quantum Langevin equations in frequency space:
\begin{equation}
\hat{\boldsymbol{D}}\left[\omega\right]=\boldsymbol{\chi}\left[\omega\right]\cdot\boldsymbol{L}\cdot\hat{\boldsymbol{D}}_{\textrm{in}}\left[\omega\right], \label{eq:langeq}
\end{equation}
where
\begin{equation}
\boldsymbol{\chi}\left[\omega\right]\equiv\left(\begin{array}{cccc}
\frac{\kappa}{2}-i\left(\omega+\Delta\right) & 0 & -iG_{-} & -iG_{+}\\
0 & \frac{\kappa}{2}-i\left(\omega-\Delta\right) & iG_{+} & iG_{-}\\
-iG_{-} & -iG_{+} & \frac{\Gamma_{m}}{2}-i\left(\omega+\delta\right) & i2\lambda e^{-i\psi}\\
iG_{+} & iG_{-} & -i2\lambda e^{i\psi}& \frac{\Gamma_{m}}{2}-i\left(\omega-\delta\right)
\end{array}\right)^{-1}. \label{eq:operX}
\end{equation}
In the experiment, we measure the output microwave spectrum through the undriven (right) cavity port. One finds the output field $\hat d_{R,\text{out}}(\omega)$ using the input-output relation $\hat{d}_{\textrm{\ensuremath{\sigma},out}}\left(\omega\right)=\hat{d}_{\sigma,\textrm{in}}\left(\omega\right)-\sqrt{\kappa_{\sigma}}\hat{d}\left(\omega\right)$. This yields
\begin{align}
\hat{d}_{R,\textrm{out}}\left(\omega\right)= \hat{d}_{R,\textrm{in}}\left(\omega\right)-\sqrt{\kappa_{R}\kappa}\left(\boldsymbol{\chi}\left[\omega\right]\right)_{11}\hat{d}_{\textrm{in}}-\sqrt{\kappa_{R}\kappa}\left(\boldsymbol{\chi}\left[\omega\right]\right)_{12}\hat{d}_{\textrm{in}}^{\dagger}
 & -\sqrt{\kappa_{R}\Gamma_{m}}\left(\boldsymbol{\chi}\left[\omega\right]\right)_{13}\hat{b}_{\textrm{in}}-\sqrt{\kappa_{R}\Gamma_{m}}\left(\boldsymbol{\chi}\left[\omega\right]\right)_{14}\hat{b}_{\textrm{in}}^{\dagger}.
\end{align}
The transmission spectrum (driven response) is
\begin{equation} \label{eq:S21}
T\left[\omega\right] = -\sqrt{\kappa_{L}\kappa_{R}}\left(\boldsymbol{\chi}\left[\omega\right]\right)_{11}.
\end{equation}
The symmetric noise spectral density is
\begin{equation}
\bar{S}_{R}\left[\omega\right] = \frac{1}{2}\int dt\left\langle \left\{ \hat{d}_{R,\text{out}}^{\dagger}\left[0\right],\hat{d}_{R,\text{out}}\left[t\right]\right\} \right\rangle e^{i\omega t}=\frac{1}{2}+\kappa_R S\left[\omega\right],
\label{eq:SR}
\end{equation}
where
\begin{eqnarray}
S\left[\omega\right] = \kappa\left|\left(\boldsymbol{\chi}\left[\omega\right]\right)_{11}\right|^{2}n_{\text{\text{c}}}^{\textrm{th}}+\kappa\left|\left(\boldsymbol{\chi}\left[\omega\right]\right)_{12}\right|^{2}\left(n_{c}^{\textrm{th}}+1\right) +\Gamma_{m}\left|\left(\boldsymbol{\chi}\left[\omega\right]\right)_{13}\right|^{2}n_{m}^{\textrm{th}}+\Gamma_{m}\left|\left(\boldsymbol{\chi}\left[\omega\right]\right)_{14}\right|^{2}\left(n_{m}^{\textrm{th}}+1\right).
\end{eqnarray}
The mechanical quadrature spectrum is
\begin{equation}
\bar{S}_{X_\phi}\left[\omega\right]=\frac{1}{2}\int dt\left\langle \left\{ \hat{X}_\phi\left(t\right),\hat{X}_\phi\left(0\right)\right\} \right\rangle e^{i\omega t},
\label{eq:SXphi}
\end{equation}
where $\hat{X}_\phi= x_{\textrm{zp}}\left(\hat{b}e^{i\phi}+\hat{b}^{\dagger}e^{-i\phi}\right)$.  The quadrature variance is given by the integral
\begin{equation}
\left\langle \hat{X}_{\phi}^{2}\right\rangle  = \int \frac{d\omega}{2\pi} \bar{S}_{X_{\phi}}(\omega).
\label{eq:nXphi}
\end{equation}

In some pump configurations, we can simplify the results. For $\delta=0$, the expressions can be simplified to
\begin{equation}
T\left[\omega\right]=-\frac{2\sqrt{\kappa_{L}\kappa_{R}}\left(\Gamma_{m}-2i\omega\right)}{4 G^2+\left[\kappa-2i\left(\omega+\Delta\right)\right]\left(\Gamma_{m}-2i\omega\right)},\label{eq:Trans}
\end{equation}
\begin{equation}
S\left(\omega\right)=\frac{4\Gamma_{m}\left[\Gamma_{m}\kappa n_{\text{\text{c}}}^{\textrm{th}}+4G_{-}^{2}n_{m}^{\textrm{th}}+4G_{+}^{2}\left(n_{m}^{th}+1\right)\right]+16\kappa n_{\text{\text{c}}}^{\textrm{th}}\omega^{2}}{\left|4 G^2+\left(\kappa+2i\omega\right)\left[\Gamma_{m}+2i\left(\omega+\Delta\right)\right]\right|^{2}},\label{eq:SR}
\end{equation}
where $G^2=G_{-}^2 - G_{+}^2$. For both $\delta=0$ and $\Delta=0$, the mechanical quadrature spectra and the quadrature variances are
\begin{equation}
\bar{S}_{X_{1,2}}\left[\omega\right] = 4 x_{zp}^2 \frac{4\kappa\left(G_{-}\mp G_{+}\right)^{2}\left(n_{\text{\text{c}}}^{\textrm{th}}+\frac{1}{2}\right)+\Gamma_{m}\left(\kappa^{2}+4\omega^{2}\right)\left(n_{m}^{\textrm{th}}+\frac{1}{2}\right)}{\left[4 G^2+\Gamma_{m}\kappa\right]^{2}+4\left(\Gamma_{m}^{2}+\kappa^{2}-8G^{2}\right)\omega^{2}+16\omega^{4}}.
\end{equation}
\begin{equation}\label{eq:nX12}
\left\langle \hat{X}_{1,2}^{2}\right\rangle = x_{zp}^2\frac{4\left(G_{-}\mp G_{+}\right)^{2}\kappa\left(2n_{\text{\text{c}}}^{\textrm{th}}+1\right)+\left[4G^{2}+\kappa\left(\kappa+\Gamma_{m}\right)\right]\Gamma_{m}\left(2n_{m}^{\textrm{th}}+1\right)}{\left(\kappa+\Gamma_{m}\right)\left(4G^{2}+\kappa\Gamma_{m}\right)},
\end{equation}
in the regime $\kappa \gg G, \Gamma_m$, Eq. (\ref{eq:nX12}) reduced to Eq. (4) in the main text.

\section{Measurement circuit}

\begin{figure}
\centering\includegraphics{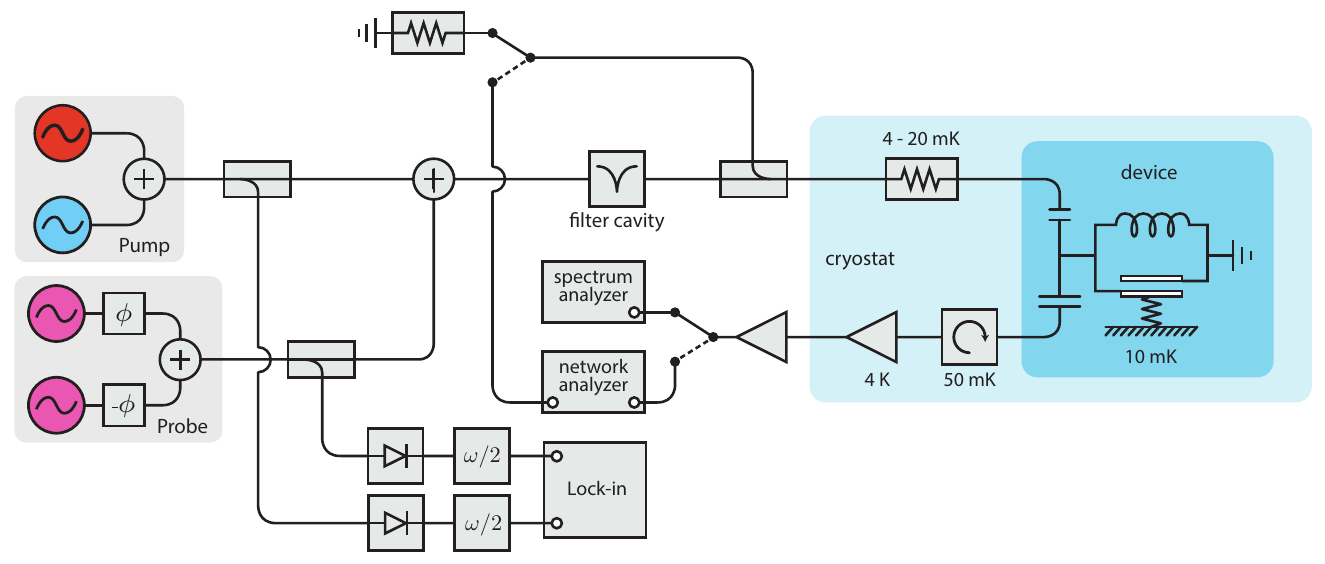}
\caption{Schematic of the measurement circuit. See text for detail.}
\label{circuit}
\end{figure}

The schematic of the measurement circuit is shown in Fig.~\ref{circuit} We cool the device with a dilution refrigerator to 10 mK. In the experiment, up to four microwave drive tones are applied to the device. Since the excess phase noise from the microwave sources at the cavity resonance can excite the cavity and degrade the squeezing. In order to avoid extra heating from the phase noise of the sources, a tunable notch filter cavity is used to provide more than 50 dB noise rejection at the cavity resonance frequency $\omega_c$. The input microwave pumps are then attenuated by about 40 dB at different temperature stages in the cryostat to dissipate the Johnson noise from higher temperature, keeping the input microwave noise at the shot noise level. The output signal passes through two cryocirculators at 50mK, then amplified by a cryogenic high electron-mobility transistor amplifier (HEMT) at 4.2 K and a low noise amplifier at room temperature for analysis. During the measurement of the noise spectrum, we continuously monitor the phase difference between the squeezing pumps and the BAE probes. The beat tones of the pumps and the probes are acquired by microwave detection diodes, then fed into the sub-harmonic circuits to halve the frequencies. The relative phase between the resulting beat tones are compared and measured by the lock-in. A computer is used to generate the error signal and feedback to the sources to keep the phase drift within half degrees.

\section{Calibrations}

\subsection{Calibration of the squeezing output spectrum}
\begin{figure}
\centering\includegraphics{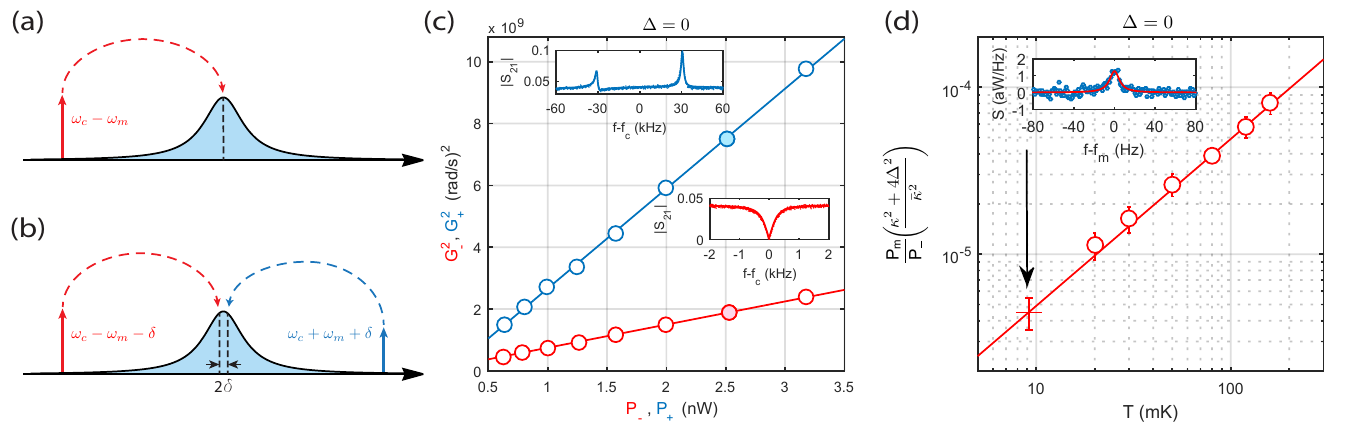}
\caption{Calibrations of the mechanical squeezing experiment. (a) Pump configuration of the enhanced optomechanical coupling $(G_-)$ calibration. (b) Pump configuration of the enhanced optomechanical coupling $(G_+)$ calibration. (c) Calibrations of the enhanced optomechanical couplings $G_{\pm}$, the inserts are the transmission spectrums corresponding to the solid circles. (d) Calibration of the normalized motional sideband power, the insert is the sideband spectrum at the base temperature.}
\label{cal_sqz}
\end{figure}

In the experiment, we spend an equal time interleaving measurement
to measure the pumped noise spectrum $\bar{S}_{\textrm{meas}}\left[\omega\right]$
and the unpumped noise spectrum $\bar{S}_{0}\left[\omega\right]$ at
the output of the measurement chain. The unpumped noise spectrum $\bar{S}_{0}\left[\omega\right]$
is the noise floor of the system which is dominated by the noise figure
of the cryogenic HEMT amplifier. The difference of the pumped and unpumped noise
spectra is related to the output noise spectrum of the optomechanical
system by
\begin{equation}
\Delta\bar{S}_{\textrm{meas}}\left[\omega\right]=\bar{S}_{\textrm{meas}}\left[\omega\right]-\bar{S}_{0}\left[\omega\right]=\mathcal{G}\left[\omega_{c}\right]\kappa_{R}\hbar\omega_{c}S\left[G_{-},G_{+},\Delta,\delta,\kappa,\Gamma_m,n_{c}^{th},n_{m}^{th},\omega\right],
\end{equation}
where $\mathcal{G}\left[\omega\right]$ is the gain of the measurement
chain around the reonance of the cavity and $\kappa_{R}$ is the coupling
rate to the output port of the device. In order
to fit the measured spectrum $\Delta\bar{S}_{\textrm{meas}}\left[\omega\right]$
to extract the detuning $\Delta$ and $\delta$, the linewidths $\kappa$ and $\Gamma_m$, occupation factors $n_{c}^{th}$and $n_{m}^{th}$,
we need an independent measurement to extract the enhanced optomechanical
coupling rate $G_{\pm}$ and the gain factor $\mathcal{G}\left[\omega_{c}\right]\kappa_{R}\hbar\omega_{c}$.

Besides the noise spectrum, we also measure the transmitted power of the drive tones at the output of the measurement chain $P_\pm$, which is related to the intracavity pump photon number by
\begin{equation}\label{Pthrured}
  P_\pm=\mathcal{G}\left[\omega_\pm\right]\kappa_R\hbar\omega_\pm\lambda\left[\omega_\pm\right]n_p^\pm,
\end{equation}
where $\lambda\left[\omega_\pm\right]$ are the correction factors due to the parasitic channel \cite{weinstein2014observation}. The square of the enhanced optomechanical couplings are related linearly to the transmitted pump powers by
\begin{equation}\label{eq:GredPthrured}
  G_\pm^2=g_0^2n_p^\pm=a_\pm\times P_\pm,
\end{equation}
where the calibration factors $a_\pm= \frac{1}{\mathcal{G}\left[\omega_\pm\right]\kappa_R\hbar\omega_\pm}\frac{g_0^2}{\lambda\left[\omega_\pm\right]}$. Therefore, we can convert the measured transmitted powers of the drive tones to the enhanced optomechanical couplings with the calibration factors $a_\pm$. In the following, we will describe the procedures to extract the calibration factors.

We start with the calibration of the enhanced optomechanical coupling $G_-$ induced by the red detuned tone. To do that, a single red detuned tone is applied at $\omega_c-\omega_m$ with transmitted power $P_-$ (Fig.~\ref{cal_sqz}a). Then, a network analyzer is used to generate a weak probe and sweep it through the center of the cavity resonance to measure the transmission spectrum of the mechanical sideband. The enhanced optomechanical coupling rate $G_-$ can be extracted by fitting the transmission spectrum with the optomechanical model (\ref{eq:S21}). By measuring the transmission spectrum with various transmitted power $P_-$ and fitting with the linear relation (\ref{eq:GredPthrured}) (the red line in Fig.~\ref{cal_sqz}c), we obtain the calibration $a_- = \left(7.49\pm0.10\right)\times10^{17} \, \textrm{rad}^2\textrm{s}^{-1}\textrm{W}^{-1}$.

A similar method can be used to calibrate the enhanced optomechanical coupling $G_+$ induced by the blue detuned tone. In this case, a blue detuned tone is placed at $\omega_c+\omega_m+\delta$ with transmitted power $P_+$, where $\delta = 2\pi\times 30 \textrm{kHz}\ll\kappa$. Since the blue detuned tone would amplify the mechanical motion and narrow the mechanical linewidth, the mechanical resonator becomes unstable when the cooperativity $C_+ = \frac{4G_+^2}{\kappa \Gamma_m}$ approaches to unity. In order to keep the mechanics stable, a constant red detuned tone is applied at $ \omega_c-\omega_m-\delta$ to damp the mechanical motion (Fig.~\ref{cal_sqz}b).
Similar to the calibration of $G_-$, we use a network analyzer to measure the transmission spectrum of the mechanical sidebands. Then we can extract the enhanced optomechanical coupling rate $G_+$ by fitting the transmission spectrum with the optomechanical model (\ref{eq:S21}). By measuring the transmission spectrum with various transmitted power $P_+$ and fitting with the linear relation (\ref{eq:GredPthrured}) (the blue line in Fig.~\ref{cal_sqz}c), we obtain the calibration $a_+ = \left(3.23\pm0.07\right)\times10^{18} \, \textrm{rad}^2\textrm{s}^{-1}\textrm{W}^{-1}$.

After calibrating the enhanced optomechanical coupling rates, in order to fit the measured noise spectrum with the optomechanical model, the last thing we need is the gain factor $\mathcal{G}\left[\omega_{c}\right]\kappa_{R}\hbar\omega_{c}$. Which can be obtained by the thermal calibration of the motional sideband noise power. To do that, a single red detuned tone is placed at $\omega_- = \omega_c-\omega_m$ with sufficiently small pump power $P_-$ such that the optomechanical damping effect is negligible $(\Gamma_{\textrm{opt}}^-=\frac{4G_-^2}{\kappa} \ll \Gamma_m)$. We then measure the noise power of the up-converted motional sideband $P_m^-$, over a range of calibrated cryostat temperature $T$ (Fig.~\ref{cal_sqz}d). Due to the weak temperature dependence of the cavity linewidth $\kappa$, we monitor the cavity linewidth at each measurement temperature. The resulting normalized sideband power is given by
\begin{equation}
\left(\frac{4\Delta^2+\kappa^2}{\bar{\kappa}^2}\right)\frac{P_m}{P_-}=b_-\left(\frac{2}{\bar{\kappa}}\right)^2\frac{k_B T}{\hbar\omega_m},
\end{equation}
where $\Delta=\omega_- - \omega_c+\omega_m$ is the detuning of the pump, which is equal to zero in this case, $\bar{\kappa}$ is the average value of the cavity linewidth over the respective temperatures and $b_-=\frac{\mathcal{G}\left[\omega_c\right]\omega_c}{\mathcal{G}\left[\omega_-\right]\omega_-}\frac{g_0^2}{\lambda\left[\omega_-\right]}$ is the thermal calibration. The linear fit in Fig.~\ref{cal_sqz}d gives $b_-=\left(2.53\pm0.07\right)\times10^5 \, \left(\textrm{rad/s}\right)^2$, which enable us to convert the normalized noise power into quanta. The gain factor is given by the ratio of the thermal calibration $b_-$ and the calibration of the enhanced optomechanical coupling $a_-$ (i.e. $\mathcal{G}\left[\omega_{c}\right]\kappa_{R}\hbar\omega_{c}=b_-/a_-$). With the calibrations discussed above, we can relate the measured noise spectrum and transmitted powers to the optomechanical model
\begin{equation}
\Delta\bar{S}_{\textrm{meas}}\left[\omega\right]=\frac{b_-}{a_-}\bar{S}\left[\sqrt{a_-\cdot P_{-}},\sqrt{a_+\cdot P_{+}},\Delta,\delta,\kappa,\Gamma_m,n_{c}^{th},n_{m}^{th},\omega\right],
\end{equation}
which enable us to extract $\Delta$, $\delta$, $\kappa$, $\Gamma_m$, $n_c^{\textrm{th}}$ and $n_m^{\textrm{th}}$ from the measured output noise spectrum.

\subsection{Calibration of the backaction evasion spectrum}
\begin{figure}
\centering\includegraphics{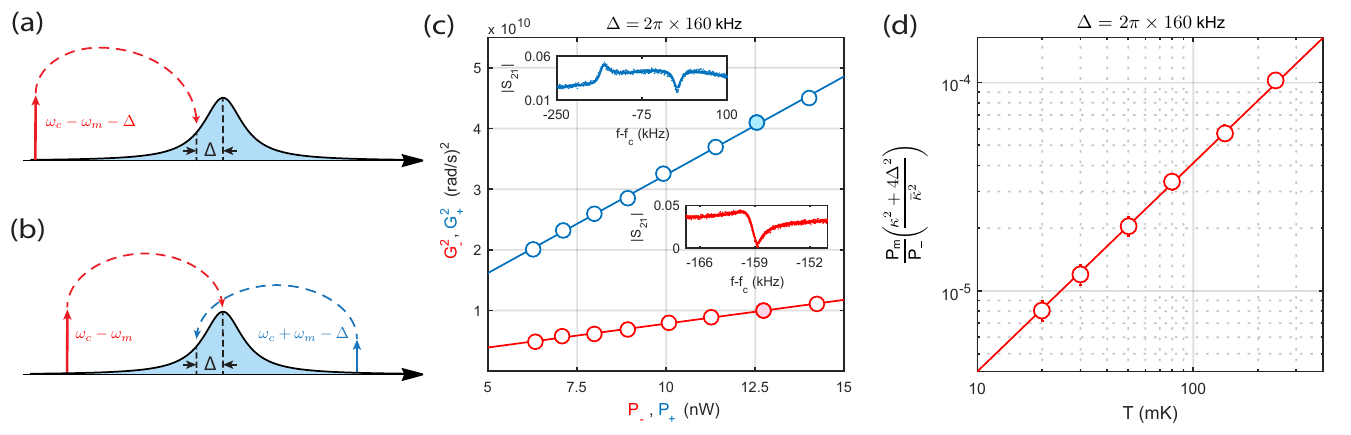}
\caption{Calibrations of the backaction evading measurement. (a) Pump configuration of the enhanced optomechanical coupling $(G_-)$ calibration. (b) Pump configuration of the enhanced optomechanical coupling $(G_+)$ calibration. (c) Calibrations of the enhanced optomechanical couplings $G_{\pm}$, the inserts are the transmission spectrums corresponding to the solid circles. (d) Calibration of the normalized motional sideband power.}
\label{cal_BAE}
\end{figure}

In our experiment, we perform an additional BAE measurement away from the cavity resonance to directly and independently measure the mechanical quadratures. Since the detuning of the BAE sideband from the cavity resonance $\Delta = 2\pi\times160\textrm{kHz}$ is comparable to the cavity linewidth, in order to precisely balance the BAE tones and correctly interpret the BAE noise spectrum, an independent calibrations of the enhanced optomechanical coupling rate $G_\pm$ and the normalized sideband power are necessary.

We follow the same procedures in the last section to calibrate the BAE measurement, the only difference is the frequency of the drive tones. Fig.~\ref{cal_BAE}a(b) is the pump configuration in calibration of the enhanced optomechanical coupling rate $G_-$($G_+$). The results of the calibrations are shown in Fig.~\ref{cal_BAE}c. From the linear fits, we get $a^{\textrm{BAE}}_- = \left(7.85\pm0.06\right)\times10^{17} \, \textrm{rad}^2\textrm{s}^{-1}\textrm{W}^{-1}$ and $a^{\textrm{BAE}}_+ = \left(3.24\pm0.03\right)\times10^{18} \, \textrm{rad}^2\textrm{s}^{-1}\textrm{W}^{-1}$. Fig.~\ref{cal_BAE}d is the calibration of the normalized motional sideband power corresponds to a single red detuned tone at $\omega_-=\omega_c-\omega_m-\Delta$. The linear fit in Fig. 3d gives $b^{\textrm{BAE}}_-=\left(2.77\pm0.04\right)\times10^5  \, \left(\textrm{rad/s}\right)^2$.

\begin{figure}
\centering\includegraphics{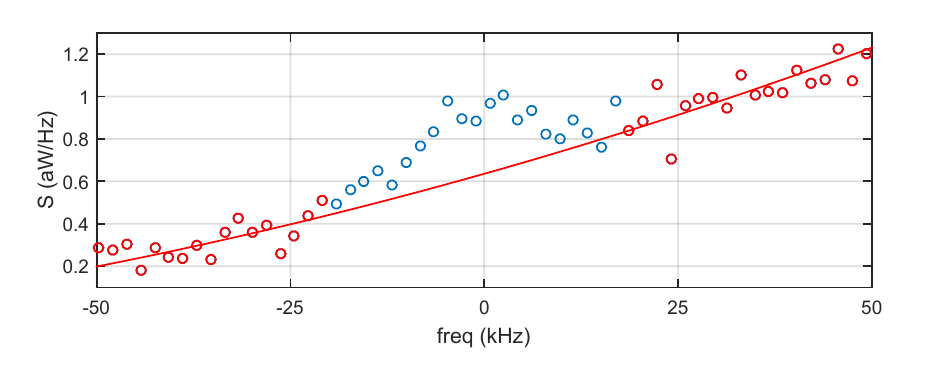}
\caption{Example of the BAE noise spectrum. The red line is a background fit with a quadratic polynomial.}
\label{fitexp}
\end{figure}

Similar to the measurement of the squeezing output spectrum, we spend an equal time interleaving measurement to measure the pumped and unpumped noise spectrum in the BAE measurement. After subtracting the unpumped noise spectrum to remove the noise floor, the noise spectrum of the BAE sideband is given by
\begin{equation}\label{eq:SRBAE}
\Delta\bar{S}_{\textrm{meas}}^{\textrm{BAE}}\left[\omega\right]=\bar{S}_{\textrm{meas}}^{\textrm{BAE}}\left[\omega\right]-\bar{S}_{0}\left[\omega\right]=\bar{S}_{c}\left[\omega\right]+\bar{S}_{\textrm{BAE}}\left[\omega\right],
\end{equation}
the first term $\bar{S}_c\left[\omega\right]$ is the noise spectrum of the microwave resonator due to the non-zero cavity occupations and the second term $\bar{S}_{\textrm{BAE}}\left[\omega\right]$ is the noise spectrum of the BAE sideband, which is given by
\begin{equation}\label{eq:SBAE}
\bar{S}_{\textrm{BAE}}\left[\omega\right]=\mathcal{G}\left[\omega_{c}\right]\kappa_{R}\hbar\omega_{c}\frac{4g_{0}^{2}}{\kappa}n_{p}\frac{\kappa}{\kappa^{2}+4\Delta^{2}}\frac{S_{X_{\phi}}\left[\omega\right]}{x_{zp}^{2}},
\end{equation}
where $S_{X_\phi}\left[\omega\right]$ is the mechanical quadrature spectrum. Because the BAE sideband is detuned from the cavity resonance with detuning comparable to the cavity linewidth, over the bandwidth of the BAE measurement, the cavity noise appears as a frequency dependent noise background. An example of the spectrum is given by Fig.~\ref{fitexp}, a quadratic polynomial is employed to fit the cavity noise background, as shown by the red curve in Fig.~\ref{fitexp}. The BAE sideband spectrum $\bar{S}_{\textrm{BAE}}\left[\omega\right]$ is given by  subtracting the noise spectrum from the fitted cavity noise background and the BAE sideband noise power is given by integrating Eq. (\ref{eq:SBAE})
\begin{equation}
P_{m}^{\textrm{BAE}}=\mathcal{G}\left[\omega_{c}\right]\kappa_{R}\hbar\omega_{c}n_{p}\frac{4g_{0}^{2}}{\kappa^{2}+4\Delta^{2}}\frac{\left\langle X_{\phi}^{2}\right\rangle }{x_{zp}^{2}}.
\end{equation}
With the thermal calibtration factor $b_-^{\textrm{BAE}}$, we can convert the normalized BAE sideband power to the quadrature variance
\begin{equation}
\frac{\left\langle X_{\phi}^{2}\right\rangle }{x_{zp}^{2}}=\frac{1}{b^{\textrm{BAE}}_{-}}\left(\frac{4\Delta^{2}+\kappa^{2}}{4}\right)\frac{P_{m}^{\text{BAE}}}{P_{-}}.
\end{equation}

\section{Analysis of the mechanical parametric effect}

\begin{figure}
\centering\includegraphics{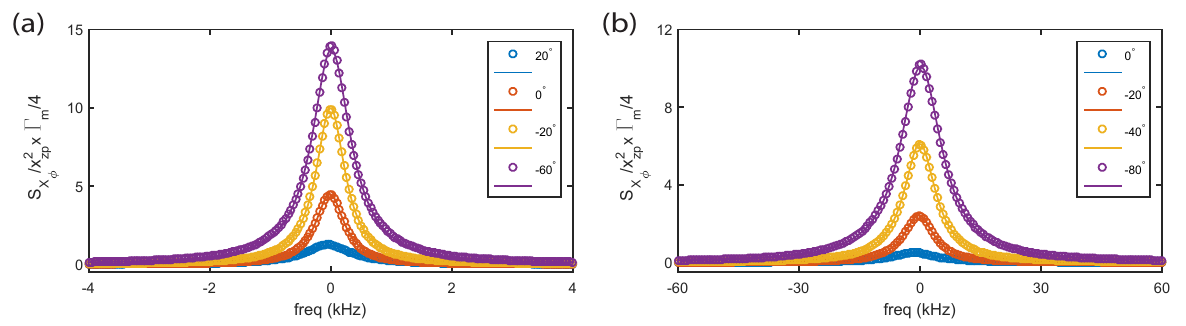}
\caption{The predicted mechanical quadrature spectra calculated with Eq. (\ref{eq:SXphi}) (circles) and the corresponding Lorentzian fits (solid curves). (a) The predicted mechanical quadrature spectra and Lorentzian fits correspond to the dashed red curve in Fig. 2c in the main text. (b) The predicted mechanical quadrature spectra and Lorentzian fits correspond to the dashed blue curve in Fig. 2c in the main text.}
\label{probepara_spec}
\end{figure}

In this section, we will describe the procedures to extract the mechanical parametric interaction from the measured quadrature linewidth data. As shown in the theory section, for given pump configurations $(\Delta,\delta,n_p^\pm)$, thermal occupations $(n_m^{\textrm{th}}, n_c^{\textrm{th}})$ and mechanical parametric interaction $(\lambda,\psi)$, the mechanical quadrature spectrum $\bar{S}_{X_\phi}[\omega]$ can be calculated by Eq. (\ref{eq:SXphi}). The mechanical quadrature linewidth is given by fitting the predicted mechanical quadrature spectrum with a Lorentzian curve, as shown in Fig.~\ref{probepara_spec}.

Using the pump configurations and thermal occupations extracted from the corresponding output spectra (Fig. 1f,g in the main text), the quadrature linewidth can be written as a function of the probe phase $\phi$, the amplitude $\lambda$ and the phase $\psi$ of the parametric drive (i.e. $\Gamma_m^{\textrm{p}}(\phi,\lambda,\psi)$). Then we can extract the mechanical parametric interaction $(\lambda,\psi)$ by fitting the measured mechanical quadrature linewidth in the BAE measurement with the function $\Gamma_m^{\textrm{p}}(\phi,\lambda,\psi)$.

\begin{figure}
\centering\includegraphics{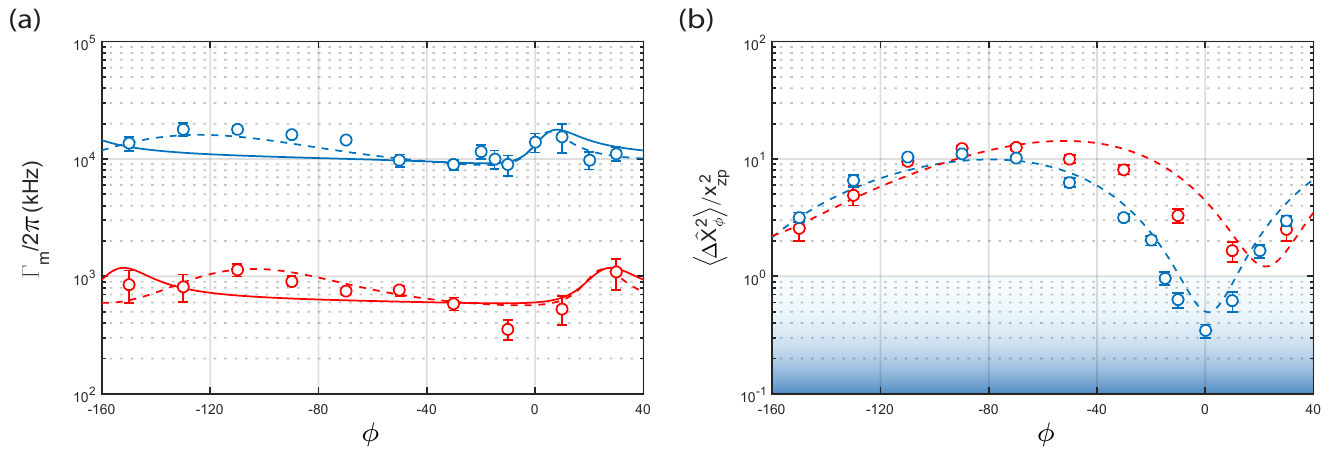}
\caption{(a) Quadrature linewidth as a function of the probe phase. The red (blue) circles are the measured quadrature linewidth of the weakly (strong) squeezed state using the BAE technique. The curves are the fits with the two-tone optomechanical model including the mechanical parametric effect. The solid curves are the fits with the assumption $\psi=\psi_0$. The dashed curves are the fits with the assumption $\psi=\phi+\psi_0$. (b) The quadrature variance as a function of probe phase. The red (blue) circles are the measured quadrature variances of the weakly (strong) squeezed state using the BAE technique. The dashed curves are the predicted quadrature variances with the assumption $\psi=\phi+\psi_0$.}
\label{pumpparafit}
\end{figure}

As discussed in the main text, we assume the phase of the parametric drive $\psi$ follows the phase of the BAE probe $\phi$ (i.e. $\psi = \psi_0 + \phi$, where $\psi_0$ is a constant phase shift). The amplitude $\lambda$ and the constant phase shift $\psi_0$ can be extracted by fitting the quadrature linewidth data with the function $\Gamma_m^{\textrm{p}}(\phi,\lambda,\psi_0+\phi)$, the fit results are shown by the dashed curves in Fig.~\ref{pumpparafit}a. From the fit, we extract $\lambda = 2\pi\times(121\pm34)$Hz, $\psi_0 = -121^\circ\pm52^\circ$ for the weakly squeezed state (red dashed curve) and $\lambda = 2\pi\times(1.3\pm0.3)$kHz, $\psi_0 = -129^\circ\pm15^\circ$ for the strong squeezed state (blue dashed curve). Under this assumption, the model captures the observed phase dependence behavior of the quadrature linewidth in the BAE measurement. With the extracted mechanical parametric drive, the corresponding quadrature variances can be calculated by Eq. (\ref{eq:nXphi}), as shown by the dashed curves in Fig.~\ref{pumpparafit}b. On the other hand, if we assume the mechanical parametric drive is induced by the squeezing pump (i.e. $\psi=\psi_0$), the function $\Gamma_m^{\textrm{p}}(\phi,\lambda,\psi_0)$ doesn't capture the observed phase dependence behavior of the mechanical quadrature linewidth, as shown by the solid curves in Fig.~\ref{pumpparafit}a. These results imply that the observed mechanical parametric interaction is induced by the BAE probes instead of the squeezing pumps..

\end{document}